\documentstyle[12pt]{article}
\input epsf.tex
\begin{document}
%\draft
\title{ Coupled electrons and pair fluctuations 
 in two dimensions: a transition to superconductivity in 
 a conserving approximation 
} 
\author{J.J. Deisz$^{a\,*}$, D.W. Hess$^b$ 
and J.W. Serene$^{a}$ \\
 \hfill  \\
%}
%\address{
$^a$Department of Physics \\
Georgetown University \\
Washington, D.C. \ \ 20057 \\
%\hfill
%}
% \address{
\hfill \\
{\em $^b$Complex Systems Theory Branch,} \\
Naval Research Laboratory, \\
Washington, D.C. \ 20375-5345 \\
}
%\date{\today}
\maketitle
\begin{abstract}
     We report on a fully self-consistent determination of a
     phase transition to a superconducting state in a conserving
     approximation.  The transition temperature calculated for
     a two-dimensional Hubbard model with an attractive interaction 
     in the fluctuation exchange approximation agrees with quantum 
     Monte Carlo calculations.  
     The temperature dependences of the superfluid density and 
     of the specific heat near T$_c$ indicate that the phase transition
     in this model of coupled collective degrees of freedom and electronic 
     degrees of freedom is consistent with neither mean-field theory, Gaussian 
     fluctuations about a mean field order parameter, nor unbinding 
     Kosterlitz-Thouless vortices. 
\end{abstract}
%\pacs{73.20.At}  
%\narrowtext 
%\widetext 

\newpage

The quasi two-dimensional nature of copper-oxide superconductors has focused
considerable attention on the role of fluctuations associated directly with 
superconductivity \cite{emery}.  In a strictly two-dimensional 
superconductor \cite{super}, thermal fluctuations of the order parameter 
preclude a finite temperature phase transition to a state with long-range 
order \cite{mermin,hohenberg}.  A Kosterlitz-Thouless transition may occur 
between two `disordered' phases, with the low temperature phase characterized 
by a finite superfluid density and correlation functions that decay 
algebraically with distance \cite{kosterlitz}. Should fluctuations be important 
for the copper-oxide superconductors, their effects would be most apparent in 
the normal state where weak three-dimensional coupling does not substantially 
influence the essentially two-dimensional character of electronic properties.  
Fluctuations of the superconducting order parameter may explain the pseudogap 
in the normal-state single-particle excitation spectrum as well as the anomalous 
temperature dependences observed for transport and thermodynamic properties in 
the normal state \cite{emery,mohit1}. 

\begin{figure}[h]
%\vspace{3.325in}
\epsfxsize=9.5truecm
\centerline{\epsffile{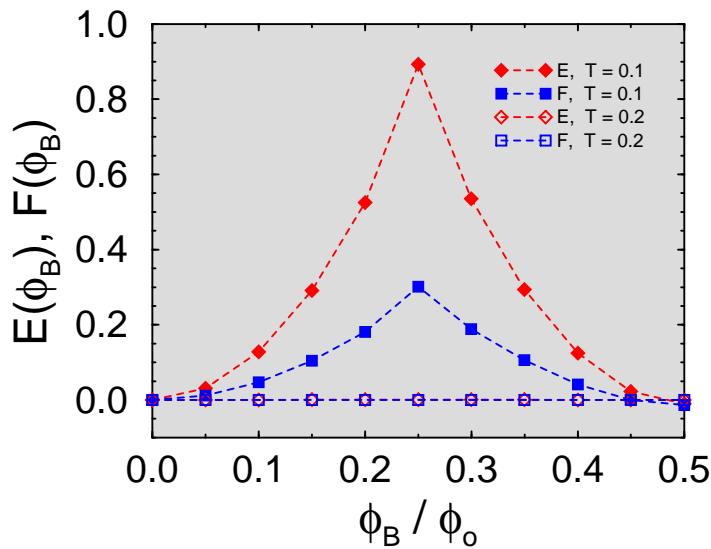}}
%\special{psfile=fig1p.ps hsize=400 vsize=260
%hscale=80 vscale=80 hoffset=-20 voffset =-43
%}
\center{
\parbox{5.5in}{
{
\caption{{ \small
Internal energy (diamonds) and free energy (squares)
as a function of the threaded flux calculated in the FEA
for a 2D Hubbard model on a $32 \times 32$ lattice
with $U=-4$  and $n=0.75$ for a temperature above (open)
and below (filled) the superconducting transition.
The energy and free energy are measured with respect to
their zero-flux values. Below the superfluid transition,
an energy barrier at $\phi_B/\phi_o = 0.25$ appears for
both $E$ and $F$ for the single period shown.
}} \label{evsphi}
}
}}
\end{figure}

To explore the role of Cooper-pair fluctuations,
we focus on the evolution of the superconducting state with decreasing 
temperature in the fluctuation exchange approximation (FEA)~\cite{bickers1}, 
a conserving approximation~\cite{baym} beyond mean field theory. 
The FEA and the Hubbard Hamiltonian~\cite{micnas} with 
an attractive interaction constitute a minimal model 
of coupled quasiparticles and s-wave Cooper-pair fluctuations.  
We report the first observation of a phase transition in the full FEA. 
The FEA superconducting phase transition occurs at a temperature in excellent 
agreement with quantum Monte Carlo (QMC) calculations. 

We probe superconductivity by calculating the internal energy $E$ and free 
energy $F$ as a function of magnetic flux $\phi_B$ threading the periodic 
lattice. Byers and Yang \cite{bandy,yang} argued that off-diagonal long 
range order makes $E(\phi_{B})$ for Cooper pairs  periodic in $\phi_B$ 
with period $\phi_0 /2 = hc/2e$. Scalapino, White, and Zhang \cite{scalapino} 
subsequently showed that the superfluid density determines the curvature of
$F(\phi_B)$ near $\phi_B = 0$, 
\begin{equation} 
F(\phi_B) \approx F(0) + {1 \over 2} \; D_s(T) \;
(\frac{\phi_B}{\phi_0})^2 + \cdots ,
\label{expandF}
\end{equation}
where $D_s(T)/2 \pi^2$ is the superfluid density in units where the hopping 
is $t = 1$ (see below).  Similarly, the energy $E(\phi_B)$ satisfies
\begin{equation}
E(\phi_B) \approx E(0) + {1 \over 2} \;
 [D_s - T \frac{dD_s}{dT}] \; (\frac{\phi_B}{\phi_0})^2
                          + \cdots, \label{expandE}
\end{equation}
which  has been used to argue for a KT transition from Monte Carlo
calculations on small lattices.  
Explicit quantum Monte Carlo 
calculations of $E(\phi_B)$ as a function of temperature for the 
two-dimensional attractive Hubbard model show a low-temperature 
superfluid state \cite{assaad1,assaad2} 
consistent with the onset of algebraic pair correlations
\cite{scalettar,moreo}.  

Fig. \ref{evsphi} illustrates one of our central results:  
$E(\phi_B)$ and $F(\phi_B)$ are flat at high temperature, but at 
low temperatures become periodic with period $\phi_o /2$ and barrier 
at $\phi_o / 4$,  indicating the formation of a low-temperature 
superconducting state \cite{energy}. 

We consider a Hubbard model on an $L \times L$ lattice with
periodic boundary conditions. In an applied magnetic flux $\phi_B$, 
the Hubbard Hamiltonian is,
\begin{eqnarray}
H(\phi_B)& = & \; -t \; \sum_{{\bf r},\sigma} \;
[ \exp [{2 \pi i {\phi_B }\over{\phi_o L}} ] \; c^{\dagger}_{{\bf r},\sigma}
c_{{\bf r} + \hat{{\bf x}},\sigma} + h.c.] \nonumber \\
&& -t \; \sum_{{\bf r},\sigma} \; [c^{\dagger}_{{\bf r},\sigma}
c_{{\bf r} + \hat{{\bf y}},\sigma} + h.c.]
+ U \sum_{{\bf r},\sigma} n_{{\bf r},\uparrow} n_{{\bf r},\downarrow},
\end{eqnarray}
where $-t$ is the nearest-neighbor hopping 
matrix element and $U$ $(<0)$ is an on-site (attractive) interaction. 

We calculate the  grand thermodynamic potential $\Omega$,
from the (self-consistent) self-energy $\Sigma$ and propagator $G$
using the Luttinger-Ward formula \cite{luttinger},
\begin{equation}
\label{grand}
\Omega \; (T,\mu,\phi_B) = - 2 \; \rm{Tr} \; [ \; \Sigma G +
\ln(-G_o^{-1} + \Sigma) \; ] + \Phi[G], \label{omega}
\end{equation}
where $\Phi [G]$ generates the skeleton diagram expansion 
for the self-energy, 
\begin{equation}
\label{phi_derive}
\Sigma({\bf k},\varepsilon_n) = {{1}\over{2}}
{{\delta\, \Phi[G]}\over{\delta \, G({\bf k},\varepsilon_n)}},
\end{equation}
which, together with Dyson's equation,
provides closed equations for $G$ and $\Sigma$.

Complete expressions for $\Phi_{FEA} [G]$ and 
$\Sigma_{FEA}[G]$ appear elsewhere\cite{serene}. 
The FEA for $\Sigma$ contains exchanged spin, density, and 
Cooper-pair fluctuations \cite{deisz2}.  The contribution from Cooper-pair 
fluctuations is $\Sigma_{\rm pp} \; ({\bf r},\tau) = 
T_{\rm pp}({\bf r},\tau)\, G(-{\bf r},-\tau)$, where 
\begin{equation}
\label{t_pp}
T_{\rm pp}({\bf q},\omega_m) = {{U^3 \; \chi_{\rm pp}^2({\bf q},
\omega_m)}\over{1 + U\chi_{\rm pp}({\bf q},\omega_m)}}
\end{equation}
is an effective non-local and retarded interaction
generated by repeated interactions of $s$-wave pairs, and 
$\chi_{\rm pp}({\bf r},\tau) = G({\bf r},\tau) \, G({\bf r},\tau)$
is the particle-particle bubble. We do not explicitly
break gauge symmetry, and hence no anomalous one-particle 
propagators enter our calculations, but off-diagonal order 
(of arbitrary spin or orbital symmetry) can still appear
spontaneously in our two-particle propagators, because 
the functional $\Phi$ is explicitly of infinite order in $U$.

The thermodynamic description of a superfluid transition requires reliable 
calculations of small differences in internal energies and free energies 
for different $\phi_B$.  These are calculated from $G$, $\Sigma$, and $\Omega$,
\begin{eqnarray}
E(\phi_B) & = & 
{1 \over 2} \; {\rm Tr} \; [ \; 
\{ 2 \epsilon_{\bf k} + \Sigma \} \; G], \label{int-energy}\\
F(\phi_B) & = & \Omega \; + \; \mu n, \label{free-energy}
\end{eqnarray}
where $\epsilon_{\bf k}$ is the bare dispersion relation, and
as in Eq. \ref{omega}, all quantities on the right-hand side of 
Eqs. (\ref{int-energy}) and (\ref{free-energy}), including $\epsilon_{\bf k}$, 
are functions of $\phi_B$.  We use a Fourier-transform-based parallel algorithm 
to solve self-consistently the FEA equations \cite{serene1,deisz1}.  
Contributions to convolution sums from high-frequency tails, essential for 
accurate calculations of thermodynamic properties, are included analytically.

Calculating $D_s(T)$ from Eq. (\ref{expandF}) using the fully self-consistent 
propagators and free energy is equivalent to obtaining $D_s(T)$ from the full 
conserving FEA current-current correlation function; hence our observation of
superconductivity is not inconsistent with previous work  that did not see 
evidence for superconductivity in an approximate FEA response 
function \cite{luo}.

\begin{figure}[h]
%\vspace{5.325in}
%\vspace{3.325in}
\epsfxsize=7.5truecm
\centerline{\epsffile{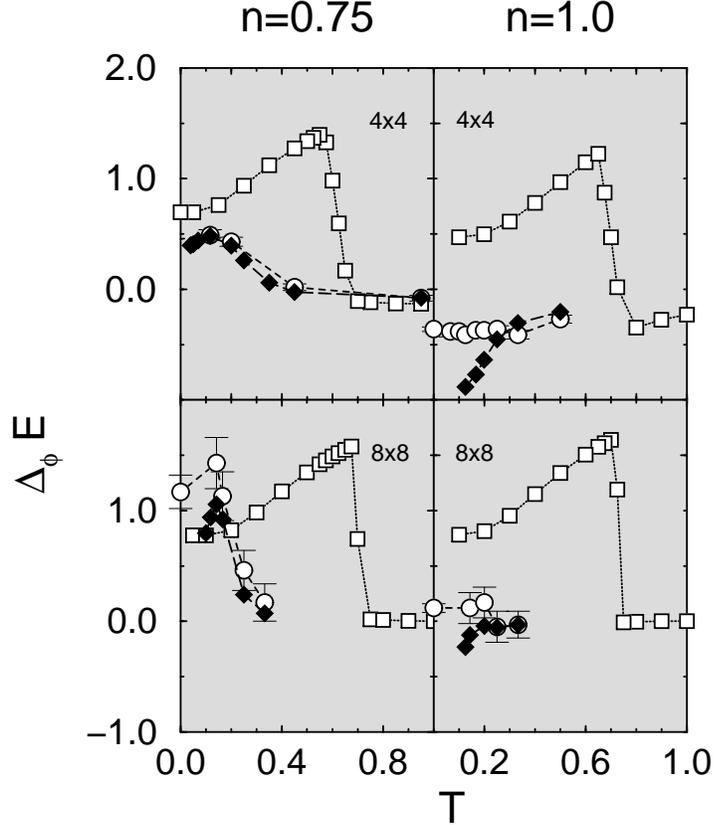}}
%\special{psfile=fig2p.ps hsize=400 vsize=260
%hscale=78 vscale=78 hoffset=-20 voffset =-73
%}
\center{
\parbox{5.5in}{
{\tiny
\caption{Temperature-dependent~energy~barriers~$\Delta_{\phi} E$,
%= E(\phi_B = \phi_o/4) - E(0)$,
for $U = -4$ calculated in the FEA (filled diamonds) shown in
comparison with the quantum Monte Carlo data of Ref. 11 (open circles)
for lattices of the same size, and $\Delta_{\phi} E$ calculated in
mean-field theory (open squares).  The FEA shows a finite
barrier (indicating superfluidity) for $n=0.75$ (left column)
and a  transition temperature well below  the mean-field theory value
of $ \simeq 0.7$ and in good agreement with the transition
temperature $ \simeq 0.1$ obtained in quantum Monte Carlo calculations
for lattices of the same size.  Like the quantum Monte Carlo
calculations, we find no evidence for superfluidity in the FEA for $n=1$.
} \label{small_lattices}
}
}}
\end{figure}

The energy shown in Fig. \ref{evsphi} was 
calculated for a $32 \times 32$ lattice with $U=-4$ and
a density $n=0.75$ \cite{energy}.
Fig. \ref{small_lattices} shows the temperature evolution of 
the energy barrier $\Delta_\phi E = E(\phi_o/4) - E(0)$ 
for small lattices in comparison with QMC results \cite{assaad1} 
and mean field theory (MFT).
Our results for $n=1.0$ show no evidence for superconductivity,
consistent with QMC calculations and theoretical expectations. 
The negative values for $\Delta_{\phi} E$ in both the QMC and
FEA and the downturn at low-$T$ in the FEA are apparently 
finite-size effects.  For $n=0.75$, the results from FEA and
QMC agree surprisingly well, but the increase with lattice size
in the maximum of $\Delta_{\phi} E$, interpreted as a signature
of a KT transition in the QMC results, is significantly smaller 
in the FEA.

\begin{figure}[h]
%\vspace{3.325in}
\epsfxsize=7.5truecm
\centerline{\epsffile{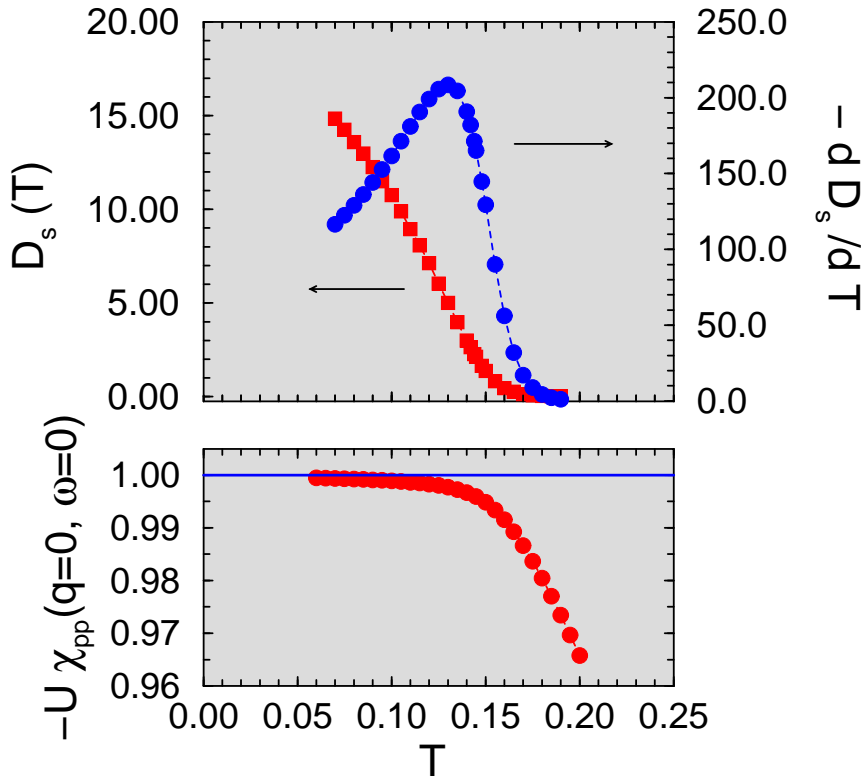}}
%\vspace{5.325in}
%\special{psfile=fig3p.ps hsize=400 vsize=260
%hscale=80 vscale=80 hoffset=-20 voffset =-73
%}
\center{
\parbox{5.5in}{
{\tiny
\caption{(top) The estimated superfluid density $D_s (T)$
(squares), and $d D_s (T)/dT$ (circles) as obtained from
the flux dependence of the free and internal energies
for a $32 \times 32$  Hubbard lattice with $U = -4$ and $n = 0.75$.
The slowly increasing $D_{s} (T)$ with decreasing temperature
does not suggest the form
 $D_{s} \propto \Theta (T_{KT} - T)$ expected for
the vortex-driven transition of Kosterlitz and Thouless.
(bottom) The onset of superfluidity is evidently
associated with a near-instability
of the particle-particle fluctuation T-matrix as signaled by
the approach of $-U \chi_{pp}({\bf q}={\bf 0},\omega_m = 0)$
to unity.
}
\label{sf_density}
}
}}
\end{figure}

The QMC calculations focussed on $\Delta_{\phi} E$, the barrier 
at $\phi_0/4$, because these calculations have significant noise
and the response is largest for this $\phi_B$.  In our calculations,
numerical noise is much smaller and we are able to use smaller
values of the flux to estimate the superfluid density and 
its temperature derivative from Eqs. (\ref{expandF}) and (\ref{expandE}).  
The top panel of Fig. (\ref{sf_density}) 
shows our estimates of $D_s(T)$ and $d D_s(T)/d T$ for a  
$32 \times 32$ lattice obtained with a flux of $\phi_B / \phi_0 = 0.1$;
errors due to higher-order terms in Eqs. (\ref{expandF}) and (\ref{expandE})
are expected to be less than $2.5\%$.

The behavior of $D_s(T)$ shown in Fig. (\ref{sf_density})
differs from expectations for both a mean-field transition
($D_s \propto T_c - T$  below $T_c$ in the thermodynamic limit)
and for a KT transition ($D_s$ is discontinuous at $T_c$ in the
thermodynamic limit).  On a $64 \times 64$ lattice, the rise in
$D_s(T)$ occurs at slightly lower temperature, but $D_s(T)$ is
otherwise very similar to the result shown for a $32 \times 32$
lattice.

The bottom panel in Fig. \ref{sf_density} shows that the superfluid 
transition occurs as $T_{\rm pp}$ approaches an instability, 
{\em i.e.} as $U \chi_{pp}$ approaches $-1$.  The T-matrix is not the 
self-consistent pair-susceptibility for the FEA 
and a near instability does not in general signal a phase transition.  
In this case, extrapolation of  
$- U \chi_{pp}({\bf q}={\bf 0},\omega_m = 0)$ for 
$T \stackrel{\normalsize <}{\normalsize \sim} 0.25$ to unity yields a 
temperature in good
agreement with that for the onset of energy-barrier formation.

\begin{figure}[h]
%\vspace{3.325in}
\epsfxsize=7.5truecm
\centerline{\epsffile{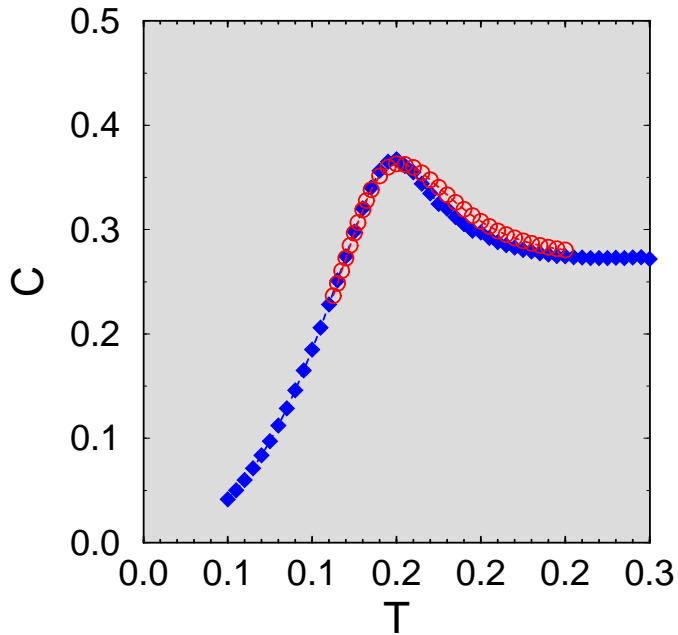}}
%\vspace{5.325in}
%\special{psfile=fig4p.ps hsize=400 vsize=200
%hscale=80 vscale=80 hoffset=-20 voffset =-73
%}
\center{
\parbox{5.5in}{
{\small
\caption{Specific heat as a function of temperature
for a $32 \times 32$ site 2D Hubbard lattice (diamonds) with 
$U = -4$ and $n=0.75$. The temperature of the peak in the specific 
heat correlates with the transition temperature.  
Also shown is the specific heat for a $64 \times 64$ lattice (open
circles) for the same density and interaction strength. 
} \label{spec-heat}
}
}}
\end{figure}

A prominent peak appears in the specific heat \cite{elsewhere} 
as shown in Fig. \ref{spec-heat}.  The peak is not consistent with MFT,
which  predicts a jump in the specific heat, nor with Gaussian
fluctuations about a mean-field order parameter.  In the latter
case fluctuations round the corners of the jump in reasonably narrow
critical region.  The specific heat is also not consistent with
expectations for a KT transition, which shows a peak above 
$T_c$ from the entropy associated with unbinding 
topological vortices.  Comparison of the specific heat and
entropy shown for $32 \times 32$ and $64 \times 64$ lattices
indicates that finite-size effects in the results shown for 
these quantities are small.

The superconducting transition we observe in the FEA is apparently
neither a simple mean-field transition nor a KT transition, unless
the FEA results for $D_s(T)$ contain unusual finite-size effects 
that die very slowly with increasing lattice size.  Our tentative
explanation is that the anomalous temperature dependence we
observe in these quantities results from the more complicated
interplay between collective degrees of freedom (order parameter
fluctuations) and electronic degrees of freedom in the FEA than in
either the BCS mean-field theory (which describes the change in
single-electron energies but ignores order-parameter fluctuations)
or descriptions based on Ginzburg-Landau functionals (which treat
fluctuations of the order parameter, but ignore their effects 
on electronic quasiparticles). Although we do not mean to suggest
that the {\em details} of the results for $D_s(T)$ or $C(T)$ from
the FEA are correct, we do think that the FEA provides a 
valuable example of how calculated properties of a superconductor can 
be modified by mutual feedback between order-parameter fluctuations and 
quasiparticle properties when the feedback is included self-consistently.

J.J.D. thanks F.F. Assaad and S.W. Pierson 
for useful conversations.
J.J.D. was supported by NSF Grant ASC-9504067.
D.W.H. is supported by the Office of Naval Research. 
This work was also supported in part by a grant of computer 
time from the Department of Defense High Performance Computing
Shared Resource Centers: Naval Research Laboratory Connection Machine 
facility CM-5; Army High Performance Computing Research Center, 
under the auspices of Army Research Office contract number 
DAAL03-89-C-0038 with the University of Minnesota.

%\begin{references}


\begin{thebibliography}{ }
\bibitem[2]{super} For our purposes, `superconductivity' is understood
to arise from short-ranged interactions; the terms `superconductivity'
and `superfluidity' will be used interchangeably.
\bibitem[17]{energy} Unless explicitly stated otherwise, all energies
are measured in terms of the nearest-neighbor 
hopping matrix element $t$.
\bibitem[24]{elsewhere} The peak in the specific heat reflects 
the strong suppression of spectral weight at and near the Fermi 
energy with decreasing temperature; we discuss this in detail 
elsewhere.
\end{thebibliography}
\end{document}